\documentclass[reprint, showpacs, showkeys, superscriptaddress, aps, pre, twocolumn, 10pt]{revtex4-1}

\usepackage[T1]{fontenc}
\usepackage[utf8]{inputenc}
\usepackage{amsmath}
\usepackage{amssymb}
\usepackage{enumitem}
\usepackage{graphicx}
\usepackage{refstyle}
\usepackage{multirow}
\usepackage[normalem]{ulem}
\newcommand\redout{\bgroup\markoverwith
{\textcolor{red}{\rule[.5ex]{5pt}{0.5pt}}}\ULon}
\usepackage[
  citecolor=blue,
  colorlinks,
  linkcolor=blue,
  urlcolor=blue,
]{hyperref}
\usepackage[dvipsnames]{xcolor}

\newcommand{\be}{\begin{equation}}
\newcommand{\ee}{\end{equation}} 
\newcommand{\bea}{\begin{eqnarray}}
\newcommand{\eea}{\end{eqnarray}}

\begin{document}
\title{Deciphering chaos in evolutionary games}
\author{Archan Mukhopadhyay}
\email{archan@iitk.ac.in}
\affiliation{
 Department of Physics,
  Indian Institute of Technology Kanpur,
  Uttar Pradesh 208016, India
}
\author{Sagar Chakraborty}
\email{sagarc@iitk.ac.in}
\affiliation{
  Department of Physics,
  Indian Institute of Technology Kanpur,
  Uttar Pradesh 208016, India
}
%
%
%
%
%
%
%
%

\date{\today}

\begin{abstract}
Discrete-time replicator map is a prototype of evolutionary selection game dynamical models that have been very successful across disciplines in rendering insights into the attainment of the equilibrium outcomes, like the Nash equilibrium and the evolutionarily stable strategy. By construction, only the fixed point solutions of the dynamics can possibly be interpreted as the aforementioned game-theoretic solution concepts. Although more complex outcomes like chaos are omnipresent in the nature, it is not known to which game-theoretic solutions they correspond. Here we construct a game-theoretic solution that is realized as the chaotic outcomes in the selection monotone game dynamic. To this end, we invoke the idea that in a population game having two-player--two-strategy one-shot interactions, it is the product of the fitness and the heterogeneity (the probability of finding two individuals playing different strategies in the infinitely large population) that is optimized over the generations of the evolutionary process.
\end{abstract}
\maketitle

\section{Motivation}
An authoritative book~\cite{cressman2003book} on evolutionary game dynamics states: \emph{``Complex dynamic behaviour can result from elementary discrete-time evolutionary processes. This is one of the main reasons dynamic evolutionary game theory deals primarily with continuous-time dynamics''}. This statement implicitly highlights the fact that the casting aside of the complex dynamical behaviour---which essentially means dynamics that don't settle down onto a stable fixed point solution---is not a rare practice in the literature of evolutionary game dynamics. This is very surprising given that the set of evolutionary processes with exclusively simple and predictable outcomes spans only a minor fraction of enormous possibilities~\cite{skyrms1992jlli,skyrms1992psa,ferriere1995tee,doebeli2014evolution}.  Probably the most omnipresent complex deterministic unpredictable behaviour is due to chaos. Of course, we are not implying that the chaos has not been explored in the context of the evolutionary game dynamics. Within the paradigms of the game theory and the theory of evolution, issues related to chaos have been presented in the context of learning~\cite{sato2002pnas,sato2003pre,sato2005physicad,sanders2018sr}, emergence of cooperation~\cite{nowak1993pnas,perc2006epl,you2017pre,krieger2020nc,chattopadhyay2020chaos}, mutation~\cite{schnabl1991pdnp,bahi2016ijb}, fictitious play~\cite{sparrow2011dgb}, imitation game of bird song~\cite{kaneko1993al}, Darwinian evolution~\cite{robertson1991jtb, robertson1996complexity}, consciousness~\cite{vandervert1995nip}, law and economics~\cite{roe1996hlr}, and language acquisition~\cite{mitchener2004prsl}. 

What, to the best of our knowledge, is lacking is an attempt to connect the chaotic behaviour with a game-theoretic concept. The reason behind this is also not hard to understand: The field of classical game theory, as in economics and other social sciences, has heavily revolved around the static equilibrium concept of the Nash equilibrium (NE)~\cite{nash1950pnas, vanDamme1991book}; researchers try to refine the notion in order to overcome the stringent requirement of rationality expected from the players of the game and to select the best equilibrium among the many possible simultaneous NEs. In the context of evolutionary biology, the equilibrium outcome of the evolutionary dynamics is expected to be an evolutionarily stable strategy (ESS)~\cite{smith1982book} that surprisingly turns out to be a refinement of the NE, although the concept of rationality is not invoked while defining the ESS. The concept of the NE becomes even more welcome to the biologists and the ethologists in the light of the fact the simple fixed point solutions of the paradigmatic replicator equation~\cite{taylor1978mb,schuster1983jtb,page2002jtb} can be tied to the ESSs (or the NEs) of the underlying game through the folk theorem of the evolutionary games~\cite{cressman2014pnas}. But, unfortunately, the non-fixed point complex solutions are not amenable to such simple convenient interpretation.

\section{Introduction}
In classical two-player--two-strategy one-shot game the NE corresponds to the strategy pair such that the strategies in the pair are the best response to each other, thereby, denying any player any gain following its unilateral strategy-deviation. The concept is easily extendable for games---which need not even be one-shot---involving many players and many strategies. Consider a normal form game with $N$ pure strategies and real payoff matrix ${\sf U}$ which is $N \times N$ dimensional. A mixed strategy, ${\bf p}$, thus, belongs to an $N-1$ dimensional simplex, $\Sigma_N$ whose vertices are the pure strategies. Using this as the underlying game, we can construct a population game~\cite{weibull1997book, hofbauer1998book,cressman2003book} between $n$ (pheno-)types that constitute fractions $x_1,x_2,\cdots,x_n$ of an infinitely large population. We represent the state of the population as a column vector, $\textbf{x}=(x_1,x_2,\cdots,x_n)^T$, that specifies a point on an $n-1$ dimensional simplex $\Sigma_n$. Here, the superscript `$T$' stands for the transpose operation. Every type can be mapped on to some strategy in $\Sigma_N$. Specifically, $i$th type---equivalently, the $i$the vertex of $\Sigma_n$---in the population game can be seen as a (possibly mixed) strategy ${\bf p}_i \in \Sigma_N$. The fitness of the $i$th type can be represented as $({{\sf \Pi} {\bf x}})_i$ where the $(i,j)$th element of the $n \times n$ payoff matrix ${\sf \Pi}$ of the population game is given by ${\bf p}_i^T{\sf U}{\bf p}_j$. In the simple case of all individual playing the same role in the population, we have a symmetric population game where a state $\hat{{\bf x}}$ is NE if $\hat{{\bf x}}^T{\sf \Pi}\hat{{\bf x}} \ge {{\bf x}}^T{\sf \Pi}\hat{{\bf x}},$ for all ${\bf x}\in\Sigma_n$. If the NE is not a pure state, then the `$\ge$' sign can be strictly replaced by an `$=$' sign. The state $\hat{\bf x}$ is furthermore an ESS of the population game if there exists a neighbourhood $\mathcal{B}_{\hat{\bf x}}$ of ${\hat{{\bf{x}}}}$ such that for all ${{{\bf{x}}}}\in \mathcal{B}_{\hat{\bf x}}\backslash \hat{\bf x}$, the following inequality holds: $\hat{{\bf x}}^{T}{\sf \Pi}{\bf{x}} >{\bf x}^{T}{\sf \Pi}{\bf {x}}$. The idea of ESS plays the central role in the evolutionary game theory since when a population is in this state, the population cannot be successfully invaded by an infinitesimal fraction of mutants with alternative strategy. Strict NE implies ESS, and ESS implies NE. This is the right place to highlight explicitly that, by definition, the NEs (and the ESSs) can be connected with only the fixed points of the replicator equation (or any such deterministic selection dynamics, if at all) that essentially is a differential equation dictating how the state of the population, ${\bf x}$, evolves in time.

It is not an exaggeration in commenting that the folk theorem has brought about a paradigm shift in the way population biologists interpret the interactions in the systems of their interest; now they can treat the individuals of the system as rational players even though in reality it is the natural selection that fashions their behaviour leading to a mathematically stable solution of the model equation (e.g., replicator dynamic). This is justified because, owing to the folk theorem, the evolutionary outcome can be predicted conveniently by finding the NE of the game modelling the inter-player competition. Naturally, when a complex chaotic dynamics is under observation, the folk theorem and the NEs are of no use. One doesn't even know how to interpret chaos in the context of the plethora of game equilibria available in the rich literature. Primarily the attempt to avoid exactly this situation resonates in the quote given in the beginning of this paper because chaos is readily witnessed in the discrete-time evolutionary processes, e.g., the one modelled by the discrete-time replicator equation in the most basic setting of two-player--two-strategy games.

Usually, the discrete-time replicator dynamic is associated with the populations having nonoverlapping generations, whereas its continuous-time version models the case of overlapping generations. However, {the} discrete-time replicator equation has been proposed for the overlapping generations case~\cite{cabrales1992jet,binmore1992book, weibull1997book} as well. While a continuous-time differential equation can be time-discretized to arrive its discrete version, a discrete-time dynamic stands on its own; e.g., the place of the discrete-time version~\cite{may1976nature} of the continuous-time logistic equation~\cite{piegton2004book} in the theory of chaos is paramount. 

\section{Replicator map and game-theoretic equilibria}
A two-player--two-strategy (one dimensional) discrete-time replicator map~\cite{vilone2011prl,pandit2018chaos,mukhopadhyay2020jtb} may be written as follows:
\begin{equation}
x^{(k+1)}=x^{(k)}+ \frac {H_{{\bf x}^{(k)}}}{2} \left[({\sf \Pi}{{\bf x}}^{(k)})_1-({\sf \Pi}{{\bf x}}^{(k)})_2\right],
\label{eq:smd_1}
\end{equation}
such that $0\le x\le1$. Here `$(k)$' denotes the time step or generation and $H_{{\bf x}^{(k)}}\equiv 2{x^{(k)}}({1-x^{(k)}})$ is the heterogeneity for a population state ${\bf x}^{(k)}=(x^{(k)},1-x^{(k)})^T$. It may be noted that $H_{{\bf x}^{(k)}}$ is the probability that two arbitrarily chosen individuals belong to two different phenotypes at the $k$th generation. In the context of one-locus--two-allele theory in the population genetics an analogous, heterozygosity, quantifies the proportion of heterozygous individuals~\citep{rice1961book}. 

Our strong motivation to work with Eq.~(\ref{eq:smd_1}) arises from the facts that (a) it is in line with the Darwinian tenet of the natural selection, i.e, only the types with fitnesses more than the average fitness of the population have positive growth rate; (b) its fixed points are related to the NE and the ESS through the folk and related theorems; and (c) it exhibits with chaotic solutions even in the simplest case of two strategies~\cite{vilone2011prl,pandit2018chaos}, specifically for the anti-coordination games like~\cite{1967_R_BS, hummert2014mbs} the leader game and the battle of sexes. We want to find the hitherto unknown game-theoretic interpretation of such chaotic solutions.

The seed of this endeavour has been sown in a recent paper~\cite{mukhopadhyay2020jtb} that shows how the concepts of the NE and the ESS may be extended to show that periodic orbits can be evolutionarily stable. All that is required is to realize that instead of fitness, heterogeneity weighted fitness, can be used to define both the (mixed) NE and the ESS---respectively redefined through $H_{\hat{{\bf x}}} \big[\hat{{\bf x}}^T{\sf \Pi}\hat{{\bf x}}\big] =  H_{\hat{{\bf x}}} \big[{{\bf x}}^T{\sf \Pi}\hat{{\bf x}}\big]$ and $H_{{\bf x}} \big[\hat{{\bf x}}^T{\sf \Pi}{{\bf x}}\big] >  H_{{{\bf x}}} \big[{{\bf x}}^T{\sf \Pi}{{\bf x}}\big]$--- using a positive definite $H_{\hat{{\bf x}}}\in(0,0.5]$. One  should note that it makes sense to exclusively work with mixed states since our goal is to comprehend the game-theoretic meaning of the non-fixed point outcomes while any pure state is a fixed point of the replicator map. 

Specifically, a periodic orbit is a heterogeneity orbit (HO($m$), which is not to be confused with the abbreviations of the homoclinic orbit or the heteroclinic orbit) and if it is asymptotically stable, it is heterogeneity stable orbit (HSO($m$)); the HO($m$) and the HSO($m$) respectively boil down to the NE and the ESS---HO($1$) and HSO($1$) respectively---when one considers fixed point as a trivial 1-period orbit. Moreover, HSO($m$) implies HO($m$) just as (mixed) ESS implies (mixed) NE. In this context it is useful to explicitly define the HO($m$) and the HSO($m$): {A sequence of states $\{\hat{\bf x}^{(k)}: \hat{ x}^{(k)} \in (0,1), k=1,2,\cdots,m\}$ where $\hat{\bf x}^{(k_1)} \ne \hat{\bf x}^{(k_2)}$ for all $k_1 \ne k_2$, of a map---$x^{(k+1)}=f(x^{(k)})$---is an HO($m$) if for all $l \in \{1,2,\cdots,m\}$,
\begin{equation}
\sum_{k=1}^{m}H_{\hat{\bf x}^{(k)}}{{\hat{\bf x}^{(l)T}}}{\sf \Pi}\hat{\bf x}^{(k)}=\sum_{k=1}^{m}H_{\hat{\bf x}^{(k)}}{{\bf x}^{T}}{\sf \Pi}\hat{\bf x}^{(k)},
\label{eq: HE3v1}
\end{equation}
for any mixed state ${\bf x}$. The sequence is furthermore an HSO($m$) if
\begin{equation}
\sum_{k=1}^{m}H_{{\bf x}^{(k)}} {{\hat{\bf x}^{(1)T}}}{\sf \Pi}{\bf{x}}^{(k)} > \sum_{k=1}^{m}H_{{\bf x}^{(k)}} {{\bf x}^{{(1)}T}}{\sf \Pi}{\bf {x}}^{(k)},
\label{eq: HSS3v1}
\end{equation}
for any trajectory $\{{\bf x}^{(k)}:{x}^{(k)} \in (0,1);\, k=1,2,\cdots,m\}$ of the map starting in an infinitesimal neighbourhood $\mathcal{B}_{{\hat{\textbf{x}}}^{(1)}}\backslash\{{\hat{\textbf{x}}}^{(1)}\}$ of $\hat{{\bf x}}^{(1)}$.} 

\section{Heterogeneity advantageous orbit }
We observe that since an HSO($m$) must obey both Eq.~(\ref{eq: HE3v1}) and Eq.~(\ref{eq: HSS3v1}) that on rearrangement yield a combined inequality defining the HSO($m$):
\begin{eqnarray}
&&\sum_{k=1}^{m}\left[H_{\hat {\bf x}^{(k)}} \left({{\hat{\bf x}^{(1)T}}}{\sf \Pi}{{\hat{\bf{x}}^{(k)}}}\right)-H_{\hat {\bf x}^{(k)}} \left({{\bf x}^{(1)T}}{\sf \Pi}{\hat{\bf {x}}}^{(k)}\right)\right]<\nonumber \\
&&\sum_{k=1}^{m}\left[H_{{\bf x}^{(k)}} \left({{\hat{\bf x}^{(1)T}}}{\sf \Pi}{{{\bf{x}}^{(k)}}}\right)-H_{{\bf x}^{(k)}} \left({{\bf x}^{(1)T}}{\sf \Pi}{{\bf {x}}}^{(k)}\right)\right],
\label{eq:HSO_1}
\end{eqnarray}
where the left hand side is identically zero. Consider that the mixed state ${{\hat{\bf x}^{(1)}}}$ when matched against the HSO($m$) equilibrium, accumulates a heterogeneity weighted fitness; and so does the mixed state ${{{\bf x}^{(1)}}}$---a state in the infinitesimal neighbourhood of the initial state of the HSO($m$). Eq.~(\ref{eq:HSO_1}) implies that the amount by which the former is more than the latter is \emph{less} than the similar difference between the accumulated heterogeneity weighted fitnesses obtained against the trajectory starting in the infinitesimal neighbourhood of the HSO($m$) equilibrium. In this sense, to be in the HSO($m$) equilibrium appears to be disadvantageous for the individuals of the population.

This motivates the question that what sequence of states is advantageous in the sense described above? Could such an orbit exist in the evolutionary dynamics? To this end first we define, what we aptly call heterogeneity advantageous orbit (HAO($m$)), as follows: A sequence of states $\{\hat{\bf x}^{(k)}: \hat{ x}^{(k)} \in (0,1), k=1,2,\cdots,m\}$, where $\hat{\bf x}^{(k_1)} \ne \hat{\bf x}^{(k_2)}$ for all $k_1 \ne k_2$, of a map---$x^{(k+1)}=f(x^{(k)})$---is an HAO($m$) if,%
\begin{eqnarray}
&&\sum_{k=1}^{m} H_{\hat{\bf x}^{(k)}} \left[{{\hat{\bf x}^{(1)T}}}{\sf \Pi}{{\hat{\bf{x}}^{(k)}}}-{{\bf x}^{(1)T}}{\sf \Pi}{\hat{\bf {x}}}^{(k)}\right]> \nonumber \\
&&\sum_{k=1}^{m} H_{{\bf x}^{(k)}} \left[{{\hat{\bf x}^{(1)T}}}{\sf \Pi}{{\bf{x}}^{(k)}}-{{\bf x}^{(1)T}}{\sf \Pi}{\bf {x}}^{(k)}\right],
\label{eq:HAO_1}
\end{eqnarray}
for any trajectory $\{{\bf x}^{(k)}:{x}^{(k)} \in (0,1);\, k=1,2,\cdots,m\}$ of the map starting in an infinitesimal neighbourhood $\mathcal{B}_{{\hat{\textbf{x}}}^{(1)}}\backslash\{{\hat{\textbf{x}}}^{(1)}\}$ of $\hat{{\bf x}}^{(1)}$. Note that in contrast with Eq.~(\ref{eq:HSO_1}), the HAO($m$) represents a set of states over a few consecutive generations such that 
on average any individual is \emph{more} efficient (in fetching heterogeneity weighted fitness) with respect to an individual in an alternate mutant-invaded state ($\{{\bf x}^{(1)},{\bf x}^{(2)},\cdots,{\bf x}^{(m)}\}$) playing against the population in the HAO($m$) ($\{\hat{\bf x}^{(1)},\hat{\bf x}^{(2)},\cdots,\hat{\bf x}^{(m)}\}$) when compared with the similar play against the population in the alternate state. In other words, the individuals of the population in the HAO($m$) enjoy an advantage compared to being in the alternate mutant-invaded population state. Obviously, the set of all possible HAOs and the set of all possible HSOs must be mutually exclusive and consequently, a stable periodic orbit---always being an HSO($m$)~\cite{mukhopadhyay2020jtb}---can never be an HAO($m$). However, it does not imply that all unstable periodic orbits are HAO($m$) because unstable periodic orbit can be HSO($m$) as well~\cite{mukhopadhyay2020jtb}. 

Chaos, observed in nature as seemingly erratic unpredictable outcomes of deterministic systems,
 is characterized in many different ways~\cite{cencini2010book,eckmann1985rmp}. For our purpose, we define~\cite{alligood1996book} a chaotic orbit as the bounded orbit that is not asymptotically periodic and has positive maximum Lyapunov exponent. Note that unlike the HO($m$), the HSO($m$), and the HAO($m$) (which are game-theoretic concepts such that they can, in principle, be defined using only the game payoff matrix when the sequences of states of interest are given), an orbit of the corresponding map and its the dynamical stability are not determined by any game-theoretic considerations but rather by the theory of dynamical systems. Thus, whether and how a chaotic orbit corresponds to the aforementioned game-theoretic concepts is an interesting question. It is however obvious that a chaotic orbit, being aperiodic and non-terminating, cannot correspond to an HO($m$). {Naturally, it is essential to relax the requirement of the vanishing of the left hand side of Eq.~(\ref{eq:HAO_1}) only while dealing with the chaotic orbits in our scheme of things.}

Further considerations bring us to the central result of this paper.
\section{Chaos and the HAO}
 \textbf{Theorem:} \emph{A heterogeneity advantageous orbit of the replicator map, Eq.~(\ref{eq:smd_1}), corresponding to the two-player--two-strategy game is either an unstable periodic orbit or a chaotic orbit.}
 
 \textbf{Proof:} Let a sequence $\{\hat{\bf x}^{(1)},\hat{\bf x}^{(2)},\cdots,\hat{\bf x}^{(m)}\}$, where $\hat{\bf x}^{(k_1)} \ne \hat{\bf x}^{(k_2)}$ for all $k_1 \ne k_2$, be an HAO($m$) of the replicator map corresponding to the two-player-two-strategy games. Therefore there exists an infinitesimal neighbourhood $\mathcal{B}_{{\hat{{{\bf x}}}}^{(1)}}$ of ${\hat{{{\bf x}}}}^{(1)}$ such that for all ${\bf x}^{(1)} \in \mathcal{B}_{{\hat{{{\bf x}}}}^{(1)}}\backslash\{{\hat{\textbf{x}}}^{(1)}\}$ (i.e., $|{{\bf x}^{(1)}-{\hat{\bf x}}^{(1)}}| \to 0^+$), Eq.~(\ref{eq:HAO_1}) is satisfied. The inequality can easily be arranged into the following form: $\left({{{x}}^{(1)}}-{\hat{{{x}}}^{(1)}}\right){ A_m}>0$, where
\begin{eqnarray}
&&{A_m}\equiv \sum_{k=1}^{m} {H_{{\bf x}^{(k)}}}\left[({\sf \Pi}\textbf{x}^{(k)})_1-({\sf \Pi}\textbf{x}^{(k)})_2\right]\nonumber \\
&&\phantom{{A_m}\equiv }-  \sum_{k=1}^{m} {H_{{\hat{\bf x}}^{(k)}}}\left[({\sf \Pi}{\hat{\textbf{x}}}^{(k)})_1-({\sf \Pi}{\hat{\textbf{x}}}^{(k)})_2\right].
\label{chaosv4}
\end{eqnarray} 
Consequently, ${ A_m}/[2\left({{{x}}^{(1)}}-{\hat{{{x}}}^{(1)}}\right)]>0$, and hence%
\begin{eqnarray}
\lim_{|{{\bf x}^{(1)}-{\hat{\bf x}}^{(1)}}| \to 0^+} \left \lvert1+\frac{{A_m}}{2({{{x}}^{(1)}}-{\hat{{{x}}}^{(1)})}}\right \rvert>1.  
\label{eq:A}
\end{eqnarray}
On further noticing that the $m$th iterate, $f^m(x)$, of the replicator map is given by,
\begin{equation}
f^{m}=x^{(1)}+ \sum_{k=1}^{m} \frac{H_{{\bf x}^{(k)}}}{2} \left[({\sf \Pi}{{\bf x}}^{(k)})_1-({\sf \Pi}{{\bf x}}^{(k)})_2\right],
\label{eq:smd_2}
\end{equation}
Eq.~(\ref{eq:A}) gets recast into
\begin{equation}
\lim_{\substack{|{{\bf x}^{(1)}-{\hat{\bf x}}^{(1)}}| \to 0^+}} \left \lvert \frac{f^{m}(x^{(1)})-{f^{m}({\hat{{{x}}}^{(1)}})}}{{{{x}}}^{(1)}-{\hat{{x}}}^{(1)}}\right \rvert >1.
\label{eq:divergence}
\end{equation}
The inequality~(\ref{eq:divergence}) implies that
\begin{equation}
\lim_{\substack{|{{\bf x}^{(1)}-{\hat{\bf x}}^{(1)}}| \to 0^+}}\left[\frac{1}{m}\ln\left \lvert \frac{f^{m}(x^{(1)})-{f^{m}({\hat{{{x}}}^{(1)}})}}{{{{x}}}^{(1)}-{\hat{{x}}}^{(1)}}\right \rvert\right]>0.
\label{eq:chaosv2}
\end{equation}
We straightaway conclude that if the HAO($m$) consists of a finite number of elements, it is an unstable periodic orbit of the map. But if the HAO($m$) consists of a nonterminating trajectory, i.e., if $m\rightarrow\infty$, then the left-hand side of Eq.~(\ref{eq:chaosv2}) is the (maximum) Lyapunov exponent if the limit exists~\cite{oseledets1968tmms,raghunathan1979ijm,eckmann1985rmp}. Consequently, we conclude that in such a case the HAO($\infty$) is a chaotic orbit. \emph{Q.E.D.}

The converse of the theorem isn't necessarily true: Recall Eq.~(\ref{eq:A}) to note that the condition of same sign of $A_m$ and $({{{x}}^{(1)}}-{\hat{{{x}}}^{(1)}})$ is only a sufficient condition for Eq.~(\ref{eq:chaosv2}) to be satisfied. In fact, if they are of different signs such that $({{{x}}^{(1)}}-{\hat{{{x}}}^{(1)}}){A_m}<-4({{{x}}^{(1)}}-{\hat{{{x}}}^{(1)}})^2$, even then Eq.~(\ref{eq:chaosv2}) is satisfied (also in the limit $m\rightarrow\infty$, given the limit exists). The condition can explicitly be written as,
\begin{eqnarray}
&&\sum_{k=1}^{m}\left[H_{\hat {\bf x}^{(k)}} \left({{\hat{\bf x}^{(1)T}}}{\sf \Pi}{{\hat{\bf{x}}^{(k)}}}\right)-H_{\hat {\bf x}^{(k)}} \left({{\bf x}^{(1)T}}{\sf \Pi}{\hat{\bf {x}}}^{(k)}\right)\right]-\nonumber \\
&&\sum_{k=1}^{m}\left[H_{{\bf x}^{(k)}} \left({{\hat{\bf x}^{(1)T}}}{\sf \Pi}{{{\bf{x}}^{(k)}}}\right)-H_{{\bf x}^{(k)}} \left({{\bf x}^{(1)T}}{\sf \Pi}{{\bf {x}}}^{(k)}\right)\right]\nonumber \\
&&<-4({{{x}}^{(1)}}-{\hat{{{x}}}^{(1)}})^2,
\label{eq:HSO_bound}
\end{eqnarray}
that clearly is not the condition of the HAO(m) but satisfies the one defining the HSO($m$) (see Eq.~(\ref{eq:HSO_1})). We observe that Eq.~(\ref{eq:HSO_bound}) is the sufficient condition for an HSO($m$) to be an unstable periodic orbit.
  \begin{figure}
	\centering
	\includegraphics[scale=0.42]{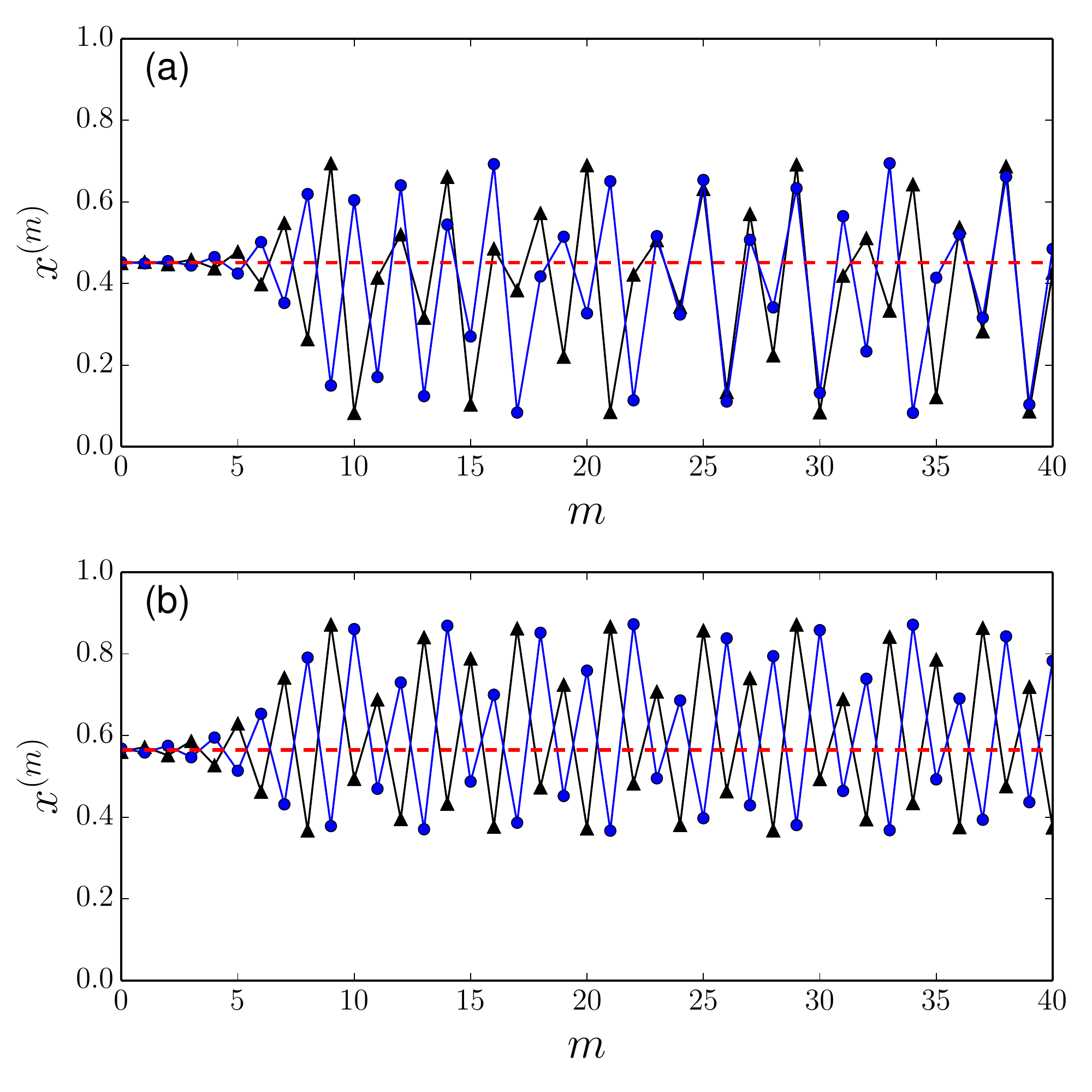} 		  
	\caption{\emph{Unattainment of ESS in chaotic two-strategy discrete-time replicator map, Eq.~(\ref{eq:smd_1}).} In subplot (a) black-triangle and blue-circle solid lines depict the chaotic time series (Lyapunov exponent=$0.39\pm0.04$) for the leader game~\cite{1967_R_BS, hummert2014mbs, pandit2018chaos} payoff matrix, $\Pi=\big(\begin{smallmatrix}
  1 & 5.34\\
  7.50 & 0
\end{smallmatrix}\big)$, starting at $x=0.450$ and $x=0.452$ respectively, both of which are in the infinitesimal neighbourhood of the ESS, $\hat{x} \approx 0.451$ (red dashed line). Similarly, in subplot (b) black-triangle and blue-circle solid lines present the chaotic time series (Lyapunov exponent=$0.11\pm0.01$) for the game of battle of sexes~\cite{hummert2014mbs, pandit2018chaos} with payoff matrix, $\Pi=\big(\begin{smallmatrix}
  1 & 6.22\\
  5.80 & 0
\end{smallmatrix}\big)$, starting at $x=0.560$ and $x=0.568$ respectively, both of which are in the infinitesimal neighbourhood of the ESS, $\hat{x} \approx 0.564$ (red dashed line). }
	\label{4.fig_1}
\end{figure} 

Note that, by construction, an HAO(1) is a mixed state that is not an ESS. (Similarly, HAO($m$), for any finite $m$, is not the extension of ESS, viz., HSO($m$).) We, however, know~\cite{weibull1997book} that ESS may be absent for a game payoff matrix or may even be non-unique, and so it is natural to wonder how the evolutionary trajectories would be in such scenarios. Moreover, even if ESS is present, it may be unattainable, e.g., in the case of continuously many infinite strategies~\cite{nowak1990aam}. In other words, mathematical existence of ESS and its practical realisation are two different aspects that are heavily dependent on the evolutionary dynamics under consideration. For example, FIG.~\ref{4.fig_1} illustrates that in the case of simple two-player--two-strategy scenario---when the payoff matrix corresponds to some anti-coordination games~\cite{kojima2007igtr}---although a mixed ESS is present, dynamically only chaotic orbits are witnessed under the discrete-time replicator map; the continuous time replicator equation~\cite{taylor1978mb}, however, leads to the ESS in this case. Such chaotic orbits naturally deserve a game theoretic explanation that is what HAO($\infty)$ may sometimes offer.

The presence of chaos in replicator equation is interesting from ecological perspective as well: Any ecological population dynamics model can be thought of as an evolutionary game with strategy dependent system parameters~\cite{vincent1988arees}. Specifically, dynamics of any $n$-strategy replicator equation can be mapped~\cite{bomze1983bc,hofbauer1998book} to the $n-1$ dimensional Lotka--Volterra equation~\cite{lotka1920jacs,volterra1926mrandl} which is widely used as a basic population dynamics model in theoretical ecology~\cite{may1973book} and mathematical biology~\cite{murray2002book1,murray2002book2}. It is easy to show that the discrete replicator map as used in this paper can similarly be mapped to the discrete-time versions of the Lotka--Volterra dynamics~\cite{hassel1976tpb,schaffer1985ecology,vincent1988arees,hastings1993arees,grafton1997mre,blackmore2001csf,bischi2010cnsns} which are known to lead to chaotic outcomes~\cite{blackmore2001csf,bischi2010cnsns} in the ecological context.

\section{Discussion and conclusion}
The connection between the stable fixed points and the ESS (a refinement of the NE) is very well known~\cite{nachbar1990ijgt,cressman2014pnas,pandit2018chaos}. This connection has been further extended~\cite{mukhopadhyay2020jtb} to show the stable periodic orbits are nothing but the HSO($m$)---a generalization of the concept of the ESS. It has been argued that in very simple reinforcement learning games (modelled by replicator dynamics and the rock-paper-scissors game)~\cite{sato2002pnas}, the players do not play NE but rather their strategies evolve chaotically over time hinting that rationality may be an unrealistic condition even in the simplest setting. Furthermore, economists have also pointed out that there is a lack of any compelling reason that real agents should play the NE~\cite{kreps1990book}; \emph{Homo Economicus} remains elusive in the real world~\cite{babichenko2017book}. Thus, dynamically---at least in the mean-field level if one considers unavoidable stochastic effects too---it is undeniable that unpredictable chaotic evolution of strategies should be present in the real world strategic interactions. This paper has presented possible evolutionary game-theoretic interpretations of such chaotic orbits (along with the unstable period orbits) arising in the replicator map.

What is crucial for such interpretations of non-convergent outcomes is to appreciate that the evolutionary game dynamics, as fashioned by the replicator map, is mathematically not about optimizing the fitness of the phenotypes; rather it is the heterogeneity weighted fitness that has to be taken into account. This is the most important implicit message of this paper. It is interesting to note that the heterogeneity can be taken as a measure of diversity in the population; it takes the maximum value when both the types of individuals are equally present and the minimum value when only one type of the individuals is present. While basic mathematical conditions of the NE and the ESS remain effectively intact even on using the heterogeneity weighted fitness, it paves way for associating evolutionary meaning to the non-fixed point outcomes in the game dynamics. In the process, we find that a chaotic attractor---that has a countably infinite number of unstable periodic orbits embedded in an uncountably infinite number of chaotic orbits---in a discrete-time replicator dynamics essentially corresponds to a collection of game-theoretic equilibria (the HSO($m$) and the HAO($m$)). 

Eq.~(\ref{eq:smd_1}) and its related forms are useful in modelling reinforcement learning~\cite{borgers1997jet}, intergenerational cultural transmission~\cite{bisin2000qje, montgomery2010aej}, and imitational behaviour~\cite{hofbauer2000jee}. In certain social scenarios, this map may be derived using the players' rational behaviour.~\cite{cressman2003book}. Our replicator equation exhibits selection monotone dynamics---a class of widely used evolutionary dynamical models~\cite{hofbauer1998book,cressman2003book} which includes replicator dynamics~\cite{taylor1978mb}, sampling best response dynamics~\cite{oyama2015te}, and stochastically perturbed best response dynamics~\cite{hofbauer2002econometrica}. We draw special attention towards the time-discrete selection monotone i-logit map that approximates both replicator map and best response map~\cite{wagner2013ps}
depending on the intensity of myopic rationality. The i-logit map leads to complex chaotic outcome~\cite{wagner2013ps}, surprisingly, for more rational players. It should be mathematically straightforward to extend the results presented thus far to the entire class of the selection monotone maps including more than two-strategy cases (see Appendices~\ref{App:A}~and~\ref{App:B}).

Ever since the focus has shifted from the existence of the static equilibrium concepts in the classical game theory of von Neumann and Morgenstern~\cite{NM1944book} towards how these equilibria are attained, the evolutionary (and similar) game dynamics have come to the fore. Since the convergent fixed-point outcomes are not an exhaustive representation of the real world, we strongly believe that---as has been the goal of this paper---one must develop new game-theoretic solution concepts that are realized as the complex dynamical outcomes, so common in nature. 
\section*{ Acknowledgments} 
The authors are thankful to Jayanta Kumar Bhattacharjee for helpful discussions. 
\section*{AIP Publishing data sharing policy} 
Data sharing is not applicable to this article as no new data were created or analyzed in this study.
\appendix
\section{Two-player--$n$-strategy game}
\label{App:A}
We consider a population game among $n$ phenotypes such that $i$th type individuals constitute fraction $x_i$  of the infinite population. Thus, the state of the population can be written as $\textbf{x}=(x_1,x_2,\cdots,x_n)^T$ that is specified by a point on an $n-1$ dimensional simplex $\Sigma_n$. A very important point to note is that, although there are now more than two types of individuals, any interaction between the individuals is still supposed to be only pairwise; i.e., as is done customarily in the replicator equation, we still have a unique payoff matrix (now $n\times n$) that specifies outcomes of any one-shot interaction since multiplayer interactions are not supposed to be occurring. This provides a hint that the heterogeneity defined in the main text for the two-player--two-strategy case should still remain pairwise in the description of the system: $H^{ij}_{{\bf x}^{(k)}}\equiv 2{x^{(k)}_i}{x^{(k)}_j}$ is the heterogeneity which for a population state ${\bf x}^{(k)}$ has been defined in a pairwise fashion. Every type contributes in $(n-1)$ different pairwise heterogeneities. Clearly, $H^{ij}_{{\bf x}^{(k)}}$ is the probability that two arbitrarily chosen individuals belong to two different phenotypes---$i$th and $j$th types---at the $k$th generation. It is hence not surprising that $H^{ij}_{{\bf x}^{(k)}}$ appear explicitly in a two-player--$n$-strategy ($n-1$ dimensional) discrete-time replicator map~\cite{borgers1997jet,hofbauer2000jee, bisin2000qje, cressman2003book,montgomery2010aej,vilone2011prl,pandit2018chaos,mukhopadhyay2020jtb} that can be written as,
\begin{equation}
x^{(k+1)}_i=x^{(k)}_i+  \sum_{\substack{j=1\\j \ne i}}^{n}\frac {H^{ij}_{{\bf x}^{(k)}}}{2} \left[({\sf \Pi}{{\bf x}}^{(k)})_i-({\sf \Pi}{{\bf x}}^{(k)})_j\right],
\label{eq:smd_1_app}
\end{equation}
such that $0\le x_i\le1$ for all $i$. Here `$(k)$' denotes the time step or generation.

It is easy to extend the concept of heterogeneity orbit (HO($m$)) and heterogeneity stable orbit (HSO($m$)) for an $n$-strategy game~\cite{mukhopadhyay2020jtb}:  {A sequence of states $\{\hat{\bf x}^{(k)}: \hat{ x}^{(k)} \in (0,1), k=1,2,\cdots,m\}$ where $\hat{\bf x}^{(k_1)} \ne \hat{\bf x}^{(k_2)}\,\forall k_1 \ne k_2$, of a map---$x_i^{(k+1)}=f(x_i^{(k)})$---is an HO($m$) if $\forall i \in \{1,2,\cdots,n\}$ and $l \in \{1,2,\cdots,m\}$, the following holds:
\begin{equation}
\sum_{k=1}^{m}\sum_{\substack{j=1\\j \ne i}}^{n}H^{ij}_{\hat{\bf x}^{(k)}}{{\hat{\bf x}^{(l)T}}_{ij}}{\sf \Pi}\hat{\bf x}^{(k)}=\sum_{k=1}^{m}\sum_{\substack{j=1\\j \ne i}}^{n}H^{ij}_{\hat{\bf x}^{(k)}}{{\bf x}^{T}_{ij}}{\sf \Pi}\hat{\bf x}^{(k)}.
\label{eq: HE3v1_app}
\end{equation}
Here ${{\hat{\bf x}}_{ij}}$ (or ${\bf x}_{ij}$) is a mixed state having same fraction of $i$th type as that of $\hat{\bf x}$ (or ${\bf x}$) but consists exclusively of $i$th and $j$th types; e.g., $\hat{\bf x}_{14}=({\hat{x}}_{1},0,0,1-{\hat{x}_{1}},0,\cdots,0)$ and ${{{\bf x}}_{31}}=(1-{{x}}_{3},0,{{x}}_{3},0,0,\cdots,0)$. The sequence is furthermore an HSO($m$) if
\begin{equation}
\sum_{k=1}^{m}\sum_{\substack{j=1\\j \ne i}}^{n}H^{ij}_{{\bf x}^{(k)}} {{\hat{\bf x}^{(1)T}}_{ij}}{\sf \Pi}{\bf{x}}^{(k)} > \sum_{k=1}^{m}\sum_{\substack{j=1\\j \ne i}}^{n}H^{ij}_{{\bf x}^{(k)}} {{\bf x}^{{(1)}T}_{ij}}{\sf \Pi}{\bf {x}}^{(k)},
\label{eq: HSS3v1_app}
\end{equation}
for any trajectory $\{{\bf x}^{(k)}:{x}^{(k)} \in (0,1);\, k=1,2,\cdots,m\}$ of the map starting in an infinitesimal neighbourhood $\mathcal{B}_{{\hat{\textbf{x}}}^{(1)}}\backslash\{{\hat{\textbf{x}}}^{(1)}\}$ of $\hat{{\bf x}}^{(1)}$.} It can be again shown~\cite{mukhopadhyay2020jtb} that a periodic orbit must be an HO($m$) and if additionally it is asymptotically stable, it is an HSO($m$) as well. Moreover, as is desirable, the HO($m$) and the HSO($m$) boil down to the NE and the ESS respectively when a fixed point is considered as a trivial 1-period orbit. Additionally, HSO($m$) implies HO($m$) just as (mixed) ESS implies (mixed) NE. 

Subsequently, in line with the main text, we define heterogeneous advantageous orbit (HAO($m$)) for two-player--$n$-strategy games as follows: {A sequence of states $\{\hat{\bf x}^{(k)}: \hat{ x}^{(k)} \in (0,1), k=1,2,\cdots,m\}$, where $\hat{\bf x}^{(k_1)} \ne \hat{\bf x}^{(k_2)}$ for all $k_1 \ne k_2$, of a map---$x_i^{(k+1)}=f(x_i^{(k)})$---is an HAO($m$) if $\forall i \in \{1,2,\cdots,n\}$,%
\begin{eqnarray}
&&\sum_{k=1}^{m}\sum_{\substack{j=1\\j \ne i}}^{n} H^{ij}_{\hat{\bf x}^{(k)}} \left[{{\hat{\bf x}^{(1)T}}_{ij}}{\sf \Pi}{{\hat{\bf{x}}^{(k)}}}-{{\bf x}^{(1)T}_{ij}}{\sf \Pi}{\hat{\bf {x}}}^{(k)}\right]> \nonumber \\
&&\sum_{k=1}^{m}\sum_{\substack{j=1\\j \ne i}}^{n} H^{ij}_{{\bf x}^{(k)}} \left[{{\hat{\bf x}^{(1)T}}_{ij}}{\sf \Pi}{{\bf{x}}^{(k)}}-{{\bf x}^{(1)T}_{ij}}{\sf \Pi}{\bf {x}}^{(k)}\right],
\label{eq:HAO_1_app}
\end{eqnarray}
for any trajectory $\{{\bf x}^{(k)}:{x}^{(k)} \in (0,1);\, k=1,2,\cdots,m\}$ of the map starting in an infinitesimal neighbourhood $\mathcal{B}_{{\hat{\textbf{x}}}^{(1)}}\backslash\{{\hat{\textbf{x}}}^{(1)}\}$ of $\hat{{\bf x}}^{(1)}$.}
It is straightforward to conclude that a stable periodic orbit can never be an HAO($m$) because all possible HAOs and the set of all possible HSOs must be mutually exclusive by definition.

We do not explicitly show the completely analogous steps of the proof given in the main text but it can be easily concluded following little inspection that a heterogeneity advantageous orbit of the replicator map, Eq.~(\ref{eq:smd_1_app}), corresponding to the two-player--$n$-strategy game is either an unstable periodic orbit or a chaotic orbit.

\section{Selection Monotone Map}
\label{App:B}
For a two-player--$n$-strategy games, the map,
	\begin{equation}
	{x_i}^{(k+1)}={x_i}^{(k)}+\phi_{i}(\textbf{x}^{(k)});\,i=1,2,\cdots,n;\label{eq:baaki}
	\end{equation}
is a selection dynamics in simplex $\Sigma_n$, if the following conditions are satisfied~\cite{cressman2003book}:
\begin{enumerate}
\item The simplex $\Sigma_n$ is forward invariant. 
\item  For all $\textbf{x}^{(k)}\in\Sigma_n$, $\sum_{i=1}^{n}\phi_{i}(\textbf{x}^{(k)})=0$.
\label{eq: 2}
\item $\phi_{i}(\textbf{x}^{(k)})$, for all $i$, is a Lipschitz continuous function on some open neighbourhood in the simplex $\Sigma_n$.
\item $\phi_{i}(\textbf{x}^{(k)})/x_i^{(k)}$, for all $i$, is continuous real-valued functions on the simplex $\Sigma_n$.
\end{enumerate}
Now, the dynamics is a \textit{monotone} selection dynamics if we impose the following condition of monotonicity: $\phi_{i}(\textbf{x}^{(k)})/x_i^{(k)} > \phi_{j}(\textbf{x}^{(k)})/x_j^{(k))}$ (whenever, $i \ne j$) if and only if $({\sf \Pi}{{\bf x}}^{(k)})_i>({\sf \Pi}{{\bf x}}^{(k)})_j$. 

The population games, having payoff or fitness being linear in the frequencies of the types of the individuals, are known as the matrix games~\cite{cressman2014pnas}. For such population the form of the selection monotone map  can easily be argued to be such that for all $j \in \{1,2,\cdots,n\}$ and $j \ne i$,
\begin{equation}
\frac{\phi_{i}(\textbf{x}^{(k)})}{x_i^{(k)}}-\frac{\phi_{j}(\textbf{x}^{(k)})}{x_j^{(k)}}=\beta (\textbf{x}^{(k)}) \Big[({\sf \Pi}{{\bf x}}^{(k)})_i-({\sf \Pi}{{\bf x}}^{(k)})_j\Big]\,,
\label{eq: msd_10}
\end{equation}
where $\beta(\textbf{x}^{(k)})$ is a positive-definite real function and should ensure that the simplex remains forward invariant. On rearranging, we get
\begin{eqnarray}
&&x_j^{(k)} \phi_{i}(\textbf{x}^{(k)})-x_i^{(k)} \phi_{j}(\textbf{x}^{(k)})= \nonumber \\
&&\frac{\beta (\textbf{x}^{(k)})}{2}   H_{{\bf x}^{(k)}}^{ij} \Big[({\sf \Pi}{{\bf x}}^{(k)})_i-({\sf \Pi}{{\bf x}}^{(k)})_j\Big]\,,
\label{eq: msd_100}
\end{eqnarray}
where the term $H_{{\bf x}^{(k)}}^{ij}=2x_i^{(k)} x_j^{(k)} $ is the pairwise heterogeneity. Now if we sum Eq.~(\ref{eq: msd_100}) for all the possible $j$ values, use condition~\ref{eq: 2} of selection monotonicity, and Eq.~(\ref{eq:baaki}), we get the general form of monotone selection dynamics for two-player--$n$-strategy matrix games as follows:
\begin{equation}
x_i^{(k+1)}=x_i^{(k)}+\sum_{\substack{j=1\\j \ne i}}^{n}\frac{\beta (\textbf{x}^{(k)})}{2}   H_{{\bf x}^{(k)}}^{ij}\left[\left({\sf \Pi}{{\bf x}}^{(k)}\right)_i-\left({\sf \Pi}{{\bf x}}^{(k)}\right)_j\right]\,. 
\label{eq: msd_20}
\end{equation} 
On comparing with Eq.~(\ref{eq:smd_1_app}), it can be easily seen that our results of the main text can readily be extended for a selection monotone map if we redefine heterogeneity by scaling with $\beta ({x}^{(k)})$ as $\beta ({x}^{(k)}) H_{{\bf x}^{(k)}}^{ij}$.

For non-matrix games (non-linear payoff function) the definition of the evolutionarily stable state (ESS) itself is somewhat problematic~\cite{cressman2014pnas}. Our main result in the main text is based on the modification of ESS beyond one-shot games. Thus, in order to extend our ideas for the selection monotone map corresponding to non-matrix game one needs to be more cautious and its careful treatment has been left for the near future. We however comment that it appears that the quantity which a chaotic orbit in it would optimize should be of the form: $\beta_{{\hat{\bf x}}^{(k)}}H^{ij}_{{\hat{\bf x}}^{(k)}} {{\hat{\bf x}^{(1)T}}_{ij}}{\bf F}\left({\sf \Pi}{{\bf{x}}^{(k)}}\right)$ where $\bf F$ is a real-valued $n$-dimensional function that is monotonically increasing with respect to its argument ${\sf \Pi}{{\bf{x}}^{(k)}}$.%

\bibliography{Mukhopadhyay_etal_manuscript.bib}
 \end{document}